\documentclass[a4paper,dvips,12pt]{article}
\usepackage{amsmath,latexsym,amssymb,epsfig,psfrag}  
\usepackage{comment}
\usepackage{cite}
\usepackage{wasysym}

\def\ltsim{\lower3pt\hbox{$\, \buildrel < \over \sim \, $}}  
\def\gtsim{\lower3pt\hbox{$\, \buildrel > \over \sim \, $}}  
\newcommand{\be}{\begin{equation}}  
\newcommand{\ee}{\end{equation}}  
\def\ga{\mathrel{\raise.3ex\hbox{$>$\kern-.75em\lower1ex\hbox{$\sim$}}}}  
\def\la{\mathrel{\raise.3ex\hbox{$<$\kern-.75em\lower1ex\hbox{$\sim$}}}}  
  
\newcommand{\myfrac}[2]{\scriptstyle#1/#2}

\newcommand{\NPB}[3]{\emph{ Nucl.~Phys.} \textbf{B#1} (#2) #3}   
\newcommand{\PLB}[3]{\emph{ Phys.~Lett.} \textbf{B#1} (#2) #3}

\makeatletter   
\@addtoreset{equation}{section}   
\makeatother

\def\simlt{\stackrel{<}{{}_\sim}}
  
\begin{document}  
  
\baselineskip=16pt   
\begin{titlepage}  
\begin{center}  
\hfill{{\bf UAB-FT-522}}

\vspace{0.5cm}  
  
\large {\sc \Large Scherk-Schwarz Supersymmetry Breaking  \\
\Large  with Radion Stabilization}

\vspace*{5mm}  
\normalsize  
  
{\bf G.~v.~Gersdorff~\footnote{gero@ifae.es}, 
M.~Quir\'os~\footnote{quiros@ifae.es}, 
A.~Riotto~\footnote{antonio.riotto@pd.infn.it}}   

\smallskip   
\medskip   
\it{~$^{1,\,2}$~Theoretical Physics Group, IFAE}\\ 
\it{E-08193 Bellaterra (Barcelona), Spain}

\smallskip   
\medskip
\it{$^2$~Instituci\'o Catalana de Recerca i Estudis Avan\c{c}ats (ICREA)}

\smallskip    
\medskip  
\it{~$^3$~Department of Physics and INFN,\\
Sezione di Padova, via Marzolo 8,
I-35131 Padova, Italy}

\vskip0.6in \end{center}  
   
\centerline{\large\bf Abstract}  

\noindent
We study the issue of radion stabilization within five-dimensional
supersymmetric theories compactified on the orbifold $S^1/\mathbb
Z_2$.  We break supersymmetry by the Scherk-Schwarz mechanism and
explain its implementation in the off-shell formulation of five
dimensional supergravity in terms of the tensor and linear compensator
multiplets.  We show that radion stabilization may be achieved by
radiative corrections in the presence of five-dimensional fields which
are quasi-localized on the boundaries through the presence of $\mathbb
Z_2$ odd mass terms.  For the mechanism to work the number of
quasi-localized fields should be greater than $2+N_V-N_h$ where $N_V$
and $N_h$ are the number of massless gauge- and hypermultiplets in
the bulk.  The radion is stabilized in a metastable Minkowski vacuum
with a lifetime much larger than cosmological time-scales. The radion
mass is in the meV range making it interesting for present and future
measurements of deviations from the gravitational inverse-square law
in the submillimeter range.

\vspace*{2mm}

\end{titlepage}  
  
\section{\sc Introduction}  \label{sec:intro}

Supersymmetry plays a crucial role in constructing consistent
high-energy physics theories and an important issue is to explain how
supersymmetry is spontaneously broken in the low-energy world. Recent
ideas on extra dimensions and the speculation that our visible
universe coincides with a four-dimensional (4D) brane living in the
bulk of the extra dimensions -- the so-called brane-world scenarios --
have given rise to new appealing possibilities regarding how to
realize supersymmetry
breaking~\cite{anton,hw,aq,nilles,mirpes,Ellis:1999dh,rs,ADPQ,gm,gp}.
In particular, non-trivial boundary conditions imposed on fields can
affect the supersymmetries of the theory. This mechanism was proposed
long ago by Scherk and Schwarz (SS)~\cite{ss} and can be interpreted
as spontaneous breaking of local supersymmetry through a Wilson line
in the supergravity completion of the theory~\cite{geromariano}.

In this paper we want to address the issue of stabilization of the
radius of the extra dimension within five-dimensional (5D)
supersymmetric theories compactified on the orbifold $S^1/\mathbb Z_2$
where supersymmetry is broken by the Scherk-Schwarz mechanism.  These
models generically exhibit a structure with a vanishing (flat)
potential for the radion, the field whose vacuum expectation value
(VEV) determines the size $R$ of the extra dimension.  Our goal is to
show that the size of the fifth dimension can be fixed in the presence
of five-dimensional fields which are quasi-localized on the boundaries
of the 5D bulk.

The wave functions of zero-modes of matter fields can be localized
towards the boundaries by adding a bulk mass-term with a non-trivial
profile in the fifth dimension~\cite{georgi}. In particular we will be
interested in 5D hypermultiplets with common odd-parity bulk masses
$M$.  Such mass terms can also appear from localized Fayet-Iliopoulos
(FI) terms corresponding to a $U(1)$ gauge group under which
hypermultiplets are charged. These FI terms, even when absent at
tree-level, can be generated radiatively~\cite{nillesgroot,bar}.
However the existence of supersymmetric odd mass terms for
hypermultiplets is more general and can even happen irrespective of
possible $U(1)$ factors in the gauge group.  Since we are interested
in an extra dimension of size $\sim$ TeV$^{-1}$, we will take $M$ to
be in the TeV range. The potential of the radion, the Casimir energy,
gets a contribution from the quasi-localized fields and the radion is
stabilized with a mass in the meV range.  In order to ensure vanishing
4D cosmological constant we introduce bulk cosmological constant and
brane tensions as counterterms, which renders the Minkowski vacuum
metastable with a stable AdS vacuum. We show that the Minkowski vacuum
turns out to be stable on cosmological times.

The paper is organized as follows. In section 2 we comment on the
Scherk-Schwarz supersymmetry breaking mechanism and its implementation
in off-shell 5D supergravity. Section 3 is devoted to the issue of
radion stabilization and the computation of the lifetime of the
metastable Minkowski vacuum. Finally, we present our conclusions and
discussion of open problems in section 4.

\section{\sc Scherk-Schwarz supersymmetry breaking}
Scherk-Schwarz (SS) supersymmetry breaking~\cite{ss} of a 5D
supersymmetric theory compactified on $S^1/\mathbb Z_2$ can be
interpreted as spontaneous breaking of 5D local
supersymmetry~\cite{geromariano}. It requires the off-shell version of
5D supergravity where an auxiliary field $\vec V_M$ gauges the
$SU(2)_R$ symmetry. This theory has been worked out in
Refs.~\cite{Zucker:1999ej,Zucker:ww} where it was shown that on top of
the minimal irreducible $40_B+40_F$ supergravity multiplet containing
the graviton ($e_M^A$), gravitino ($\psi_M$), graviphoton ($A_M$) and
auxiliary fields, an aditional $8_B+8_F$ supermultiplet is needed
which plays the role of a compensator multiplet and is purely
auxiliary.  The most convenient choice for the compensator multiplet
in order to discuss SS supersymmetry breaking is the so-called tensor
multiplet consisting of an $SU(2)_R$ triplet $\vec Y$, a fermionic
$SU(2)_R$ doublet $\rho$, a three-form tensor field $B_{MNP}$ and a
real scalar $N$.  The minimal multiplet contains a real auxiliary
scalar $C$ which enters in the action only through the term
\begin{equation}
C\left(1
-(\vec Y\vec Y)^{\frac{1}{2}}\right).
\end{equation}
The field $\vec Y$ is then constrained\footnote{A similar fermionic
Lagrange multiplier fixes the spinor $\rho$ to zero. Both conditions
receive corrections in the presence of additional vector- and
hypermultiplets in the bulk.}  to obey $\vec{Y}\vec{Y}=1$ which breaks
$SU(2)_R$ spontaneously down to $U(1)_R$. A convenient gauge fixing is
provided by
\begin{equation}
\vec Y=\left(0,1,0\right),
\label{gaugefixing}
\end{equation}
such that the surviving $U(1)_R$ is generated by $\sigma_2$. After
this gauge fixing the gravitino kinetic term is only covariant with
respect to~$U(1)_R$ according to~\footnote{To simplify the notation we
are hereafter removing the superscript from $V_M^2$ and define
$V_M\equiv V_M^2$.}
\begin{equation}
\mathcal D_M=D_M- i\frac{\sigma_2}{2}V_M, 
\end{equation}
where $D_M$ is the usual covariant derivative containg spin- and other
possible gauge connections. In the background $V_5$ the gravitino
receives a mass and supersymmetry is spontaneously broken.  The same
holds true for other $SU(2)_R$ non-singlets such as hyperscalars and
gauginos which are doublets and possible matter in tensor-multiplets
which contain bosonic $SU(2)_R$-triplets. SS supersymmetry breaking
has also been studied in the context of gauged (AdS-)
supergravity~\cite{Lalak:2002kx,Bagger:2003fy}.  In the off-shell
version we are using one can gauge the remaining $U(1)_R$-symmetry
straightforwardly~\cite{Zucker:ww}.  The equations of motion then
yield that $V_M$ is proportional to $g A_M$ where $g$ is the AdS gauge
coupling.  A non-zero VEV for the fifth component of the graviphoton
$A_5$ thus induces a VEV for the auxiliary field $V_5$. Our
(tree-level) supergravity action corresponds to ungauged supergravity
with $g=0$.

Let us see under which circumstances the VEV of $V_5$ is either
undetermined at tree-level or fixed by an explicit source.  The former
case is realized by integrating out the three-form-field of the
compensator tensor multiplet~\cite{geromarianotoni}
(section~\ref{tensor}), while supersymmetry-breaking can be realized
at tree level by switching to a dual theory in terms of the so-called
linear multiplet~\cite{Rattazzi:2003rj} (section~\ref{linear}). Both
theories are equivalent locally and only differ by their global
properties although we will also show that there exists a
linear-multiplet formulation which is globally equivalent to the
tensor one.

\subsection{\sc Tensor multiplet formalism}
\label{tensor}

It turns out that after integrating out all auxiliary fields except
$V_M$ and $B_{MNP}$, the Lagrangian containing the latter fields is
\be
\mathcal{L}=-V_M W^M+W^MW_M, 
\label{LBV}
\ee
where the field strength of the three-form tensor field is given by
\be
W^M=\frac{1}{12}\epsilon^{MNPQR}\partial_N B_{PQR}- J^M
\label{Wdef}
\ee
and the gravitino current $J^M$ is defined to be 
\be
J^M=-\frac{1}{4}\overline\psi_P \gamma^{PMQ} \sigma^2 \psi_Q.
\label{gravitinocurrent}
\ee
Note that Eq.~(\ref{LBV}) is covariant with respect to~$U(1)_R$
transformations once the full Lagrangian is considered, since the
first term makes part of the covariant gravitino derivative.  
In order to perform the integral over $B$ properly~\footnote{The
equations of motion for $V$ and $B$ give $W=0$ and $dV=0$. The former
equation gives a global condition on the gravitino current that is
lost when plugging both equations into Eq.~(\ref{LBV}), giving exactly
zero (see discussion in Ref.~\cite{Rattazzi:2003rj}). Our approach
preserves this condition and still allows to eliminate $B$.} we
linearize the $B$-dependent terms by introducing an auxiliary field
$X_M$
\be
\mathcal{L}=-V_M W^M+W^MX_M-\frac{1}{4}X_MX^M.
\label{LBVX}
\ee
Varying with respect to $B$ we find the equation
\be
C_{MN}\equiv 2\partial_{[M}(V-X)_{N]}=0.
\ee
We implement this condition by a functional delta in the path-integral
\be
\delta[C_{MN}]=\lim_{\xi\to 0}\exp\left(-i\frac{1}{2\xi}C_{MN}C^{MN}\right)
\label{functionaldelta}
\ee
where we have omitted a $\xi$-dependent but irrelevant normalization
factor~\footnote{This is a well known technique in QED, where the
Landau gauge $\partial^MA_M=0$ can be fixed by the functional
$\delta[\partial^MA_M]$ giving $\mathcal L_{\rm
g.f.}=-\frac{1}{2\xi}(\partial^MA_M)^2$. The Landau gauge then
corresponds to $\xi=0$.}.  We thus get the Lagrangian
\be
\mathcal{L}=-\frac{1}{2\xi}C_{MN}C^{MN}+(V_M-X_M)J^M -\frac{1}{4}X_MX^M
\label{LVX}
\ee
where the theory is understood in the limit $\xi\to 0$.  We now make a
shift of variables in the path integral $V_M\to V_M+X_M$ and finally
integrate over $X_M$ to obtain:
\be
-\frac{1}{2\xi}V_{MN}V^{MN}+V_MJ^M
\label{LVJ}
\ee
where $V_{MN}=2\partial_{[M}V_{N]}$ is the field strength of $V$.
Eq.~(\ref{LVJ}) fixes $V_M$ to be a closed form in the limit $\xi\to0$
and thus it can have a non-zero flux:
\be
\frac{1}{2\pi}\oint dx^5\langle V_5\rangle\equiv 2 \omega.
\label{wilsonflux}
\ee
This nontrivial flux can be removed by a nonperiodic
gauge-transformation thereby transmuting the explicit mass terms for
$SU(2)_R$ non-singlets into Scherk-Schwarz boundary conditions for
those fields.

In short in the tensor multiplet formalism, after integrating out all
auxiliary fields at the tree-level, one gets that $V$ is a closed but
not necessarily an exact form. In spaces which allow for a non-trivial
cohomology this means that $V$ can have a non-vanishing but (at
tree-level) undetermined Wilson flux as in
Eq.~(\ref{wilsonflux}). Fixing $\langle V_5\rangle$ (i.e.~the Wilson
flux) should then be done by higher-loop corrections. It is important
to realize that this procedure does not violate non-renormalization
theorems: these imply that if there is a classical supersymmetric
minimum for an auxiliary field its energy remains zero after including
quantum corrections. Furthermore any other dynamically generated
minimum should have negative energy and would be outside the validity
regime of perturbation theory. One concludes that a classical
supersymmetric minimum is not renormalized.  The fundamental
difference in our case is that there is no such classical minimum and
the tree-level potential for $V_5$ is exactly flat.  We therefore
conclude that our perturbative analysis should be
valid~\footnote{Although the field $V$ is not propagating, it
nevertheless makes sense to compute its VEV $\langle V_5\rangle$ in
the quantum theory if it is undetermined at the clasical level. Note
that we do not rely on any dynamical evolution of a classical minimum
to a quantum one.}.

To conclude we will comment on the Lagrangian (\ref{LVJ}) that is
singular in the limit $\xi\to 0$. The general strategy is then to
first compute physical observables as functions of $\xi$ and then take
the $\xi\to 0$ limit. In particular any observables obtained by
exchanging $V_M$ as internal lines will vanish in that limit. Moreover
quantities where $V_M$ are external lines, as the effective potential
$V_{\rm eff}(V_5)$, will be insensitive to that limit.

\subsection{\sc Linear multiplet formalism}
\label{linear}

In this section we will present a formulation which uses the so-called
linear multiplet as the compensator for the $SU(2)_R$ symmetry.  It
consists of the fields $(\vec Y,\rho,N, W_A)$ where $W$ obeys the
constraint $\partial_M(W^M+J^M)=0$ or in the language of differential
forms
\be
d\!*\!(W+J)=0
\label{localconstraint}
\ee
in order to ensure closure of the supersymmetry algebra. Of course
the tensor multiplet is a special case of the linear multiplet where
the constraint is explicitely solved by a three-form potential for
$W$. 
The difference of the tensor multiplet and the linear multiplet
lies in the additional global constraint on the former
\be
\intop\!*(W+J)=0.
\label{globalconstraint}
\ee
where the integral goes over any four-cycle.  This constraint must
be implemented if we insist on a completely equivalent description of
the tensor multiplet in terms of the linear multiplet.  A caveat of
the linear multiplet formalism is that the full gauge invariant
off-shell action cannot be written in terms of the linear multiplet
(see the analogous result in 4D $N=2$
supergravity~\cite{deWit:1982na}).  However, after gauge-fixing the
action for the tensor-multiplet only depends on $dB$ and consequently
one can consider the field strength as the independent variable and
ensure constraints (\ref{localconstraint}) and
(\ref{globalconstraint}) by introducing a Lagrange multiplier.

Before doing so let us study an intriguing possible interaction of the
linear multiplet. Consider a generic vector field $E_M$ with vanishing
field strength $dE=0$. Then we can define a Maxwell-multiplet
\be
\mathbb E\equiv(E_M,M=0,\Omega=0,\vec X=0).
\label{fluxmult}
\ee
where all other components except $E_M$ (i.e.\ the scalar $M$, the
gaugino $\Omega$ and the auxiliary triplet $\vec X$) are set to zero.
This configuration is left invariant under local supersymmetry
transformations with parameter $\epsilon(x^M)$, since $\delta_\epsilon
\mathbb E$ only depends on $E_M$ through its field strength $dE$.
This multiplet has no physical degrees of freedom, however on
non-simply connected spaces it can have a flux $\int E_Mdx^M$ which
might have some physical impact. Let us therefore refer to $\mathbb E$
as the {\em flux multiplet}. Under the $\mathbb Z_2$ compactification
with $E_5$ even, the flux-multiplet reduces to the constant multiplet
$(\int E_5 dx^5,0,\dots)$ which is known to be locally
supersymmetric~\footnote{Would $E_\mu$ be considered as even we would
get a 4D flux multiplet.}.  Using the locally supersymmetric coupling
of a tensor- to a Maxwell-multiplet~\cite{Zucker:ww}
we find the Lagrangian
\be
\left(W^M+J^M
\right)E_M-\frac{1}{2\xi}\left(E_{MN}\right)^2
\label{Wbrane}
\ee
where we already performed the gauge-fixing (\ref{gaugefixing}).  The
last term is again the functional delta already encountered in
Eq.~(\ref{functionaldelta}). It ensures the constraint $dE=0$ in the
limit $\xi\to0$. Let us stress that this term is needed to ensure
supersymmetry since for $dE\neq 0$ the multiplet Eq.~(\ref{fluxmult})
is not supersymmetric. However we expect a violation of supersymmetry
to $\mathcal O(\xi)$.

We will use the interaction Eq.~(\ref{Wbrane}) to enforce the local
and global constraints (\ref{localconstraint}) and
(\ref{globalconstraint}).  Indeed, varying Eq.~(\ref{Wbrane}) with
respect to~$E$ we find (in the language of differential forms)
\be
*(W+J)=\frac{1}{\xi}d\!*\!dE.
\ee
We see that now the local and global constraints on $W$ are
implemented.  We expect therefore that the theory
defined by the Lagrangian
\be
\mathcal{L}=-V_M W^M+W_MW^M+
\left(W^M+J^M
\right)E_M-\frac{1}{2\xi}\left(E_{MN}\right)^2
\label{final}
\ee
is equivalent to the one described in section \ref{tensor}.
Indeed, after integrating out $W_M$ the Lagrangian coincides with that
obtained in the tensor multiplet formalism, Eq.~(\ref{LVX}), after the
identification $X_M=V_M-E_M$. From there on we would get a non-zero
(tree-level undetermined) flux for $\langle V_5\rangle$. This exhibits
the global equivalence of both linear and tensor formalisms.

If one does not insist on global equivalence of the linear and tensor
multiplet formalisms, it is possible to fix the VEV of $V_5$ at tree
level using an independent source of supersymmetry breaking that will
play the role of the superpotential $W$ in the low energy effective
theory. In Refs.~\cite{Bagger:2001ep,Rattazzi:2003rj} this was
achieved by attaching this superpotential to the branes at $y=0,\pi
L$.  In our formalism we thus choose the flux multiplet with
$E_M=\delta_M+\partial_M \Lambda$ where $\Lambda$ is a scalar Lagrange
multiplier field, $\delta_\mu=0$ and
\be
\delta_5=\omega_0\delta(x^5)+\omega_\pi \delta(x^5-\pi R).
\label{delta5}
\ee
Obviously
$dE\equiv 0$, so we can write Eq.~(\ref{Wbrane}) without the
$\xi$-terms:
\be
\left(W^M+
J^M
\right)(\delta_M+\partial_M \Lambda)
\label{Wbrane2}
\ee
Varying with respect to $\Lambda$ enforces only the local constraint
Eq.~(\ref{localconstraint}) but not the global one
Eq.~(\ref{globalconstraint}).  We can even generalize the source term
by substituting $\delta$ by any closed but fixed one-form such as the
constant one $\omega dx^5$. According to our general analysis such a
term is supersymmetric, so we conclude that there is nothing special
about the orbifold and we can use this particular formalism to
implement supersymmetry breaking on the circle~\footnote{One should also worry
that such a term is not breaking general coordinate invariance. In
fact a fixed closed one form $E_M$ is not the same in any frame but
differs by a gauge transformation:
$$ \delta_{\xi}E_M=-\partial_M \xi^N E_N -
\xi^N\partial_NE_M=-\partial_M(\xi^N E_N).  
$$
where in the last step we have used the condition $dE=0$.
This tells us that $E_M$ is the same in all frames modulo
gauge transformations, which will drop out of the action.
}.
Our final off-shell Lagrangian is thus the sum of Eqs.~(\ref{LBV}) and 
(\ref{Wbrane2}):
\be
\mathcal L=-V_M  W^M+W^MW_M+\left(W^M+
J^M
\right)(\delta_M+\partial_M \Lambda)
\label{final2}
\ee
This leads immediately to the on-shell Lagrangian
\be
J^M(\delta_M+\partial_M \Lambda).
\label{final3}
\ee
We conclude that it is now the flux of $\delta$ which triggers
supersymmetry breaking~\footnote{Note that $dX$ does not contribute to
the flux.}. 

Notice that the gravitino mass term from Eq.~(\ref{final3}),
$\delta_5J^5$, where $\delta_5$ is defined in Eq.~(\ref{delta5}) does
not quite agree with the similar one $\delta_5J^5_{\rm RSS}$, with
$J^5_{\rm RSS}=\frac{1}{4} \overline
\psi_\mu\gamma^{\mu\nu}\sigma^1\psi_\nu$, used in Eq.~(5.6) of
Ref.~\cite{Rattazzi:2003rj}. In fact using 4D Majorana notation
($\psi_\mu^1$ even and $\psi_\mu^2$ odd) it is easy to see that the
difference $J^5-J^5_{\rm RSS}$ is proportional to $\overline{
\psi^2}_\mu\gamma^{\mu\nu}\psi_\nu^2$, i.e.~to an odd$\times$odd term.
These terms can not be disregarded since odd fermion fields may be
discontinous across the brane and behave like the step function
$\epsilon(y)$ close to $y=0$. In fact it is easy to show that
$\epsilon^{2n}(y)\delta(y)=\frac{1}{2n+1}\delta(y)$ and thus even
powers of odd fields may couple to the
brane~\cite{Meissner:2002dg,Bagger:2002rw,Delgado:2002xf,
Lalak:2003fu,Choi:2003kr}.  Consequently the mass eigenvalues and
eigenfunctions deduced from the different mass terms are not the same:
the term $\delta_5 J^5$ results in a shift $\frac{1}{\pi
R}(\omega_0+\omega_\pi)$ with respect to the KK masses $n/R$, while in
the case of the mass term $\delta_5J^5_{\rm RSS}$ this shift is given
by $\frac{1}{\pi R}(\arctan \tanh \omega_o+\arctan\tanh
\omega_\pi)$~\cite{Delgado:2002xf,Choi:2003kr}.

Note that an additional explicit mass term $\delta_M(W^M+J^M)$ in
Eq.~(\ref{final}) would not have any effect since it can be absorbed
into $E_M$ by a shift.  In any case, whether we fix the flux
radiatively or at tree level, the breaking corresponds to the
Scherk-Schwarz mechanism since the Wilson line can be removed in all
cases by means of a non-periodic gauge transformation yielding
non-trivial boundary conditions for $SU(2)_R$ non-singlets.

Let us finally comment on the low energy effective theory. A
straightforward compactification maintaining only zero modes of the
Lagrangian (\ref{final2}) gives rise to a no-scale model with constant
superpotential $\omega_0+\omega_\pi$. However at one-loop heavy KK
modes need to be properly integrated out and the Casimir energy that
will be necessary to fix the radion VEV spoils the no-scale structure.
These corrections give rise to modifications of the K\"ahler potential
as calculated in Refs.~\cite{Buchbinder:2003qu,Rattazzi:2003rj} and in
general also to terms which involve higher superspace derivatives of
the radion superfield and are not contained in the standard
parametrization of 4D supergravity in terms of K\"ahler- and
superpotential.  These terms become especially important if
supersymmetry breaking is not fixed at tree level and therefore have
to be included in order to get a consistent low energy effective
theory.

\section{\sc Radion stabilization}

In this section we will consider radion stabilization using the
Casimir energy.  We parametrize the 5D metric in the Einstein frame
as~\cite{Appelquist:1982zs}
\be ds^2=G_{MN}dx^M dx^N\equiv\phi^{-\frac{1}{3}}g_{\mu\nu} dx^\mu
dx^\nu+\phi^{\frac{2}{3}} dy^2.
\label{metrica}
\ee
where $y=x^5$ goes from $0$ to $L$. The radion field, whose VEV
determines the size of the extra dimension is $\phi^{\frac{1}{3}}$ and
the physical radius is given by
\be 
R=\frac{1}{2 \pi}\oint dy \sqrt{\langle G_{55}\rangle} = 
\langle\phi\rangle^\frac{1}{3} L. 
\ee 
The length scale $L$ is unphysical and completely arbitrary. It will
drop out once the VEV of the radion is fixed and the effective 4D theory
will only depend on $R$.

In order to achieve zero four-dimensional cosmological constant we
will introduce bulk cosmological constant and brane tensions as
possible counterterms.  This corresponds to AdS$_5$ supergravity,
although the AdS gauge coupling $g$ (as well as the brane tensions)
are really one loop counterterms and there is no tree level
warping~\footnote{To fine tune the four dimensional cosmological
constant to zero one might not introduce any bulk cosmological
constant and use only brane tensions. However this would be in
conflict with local 5D supersymmetry since the absolute value of the
brane tensions are bounded by the AdS gauge coupling
$g$~\cite{Bagger:2002rw,Bagger:2003vc} (see
also~\cite{Lalak:2003fu}). This is explained in more detail below.}.
The relevant counterterms are given by
\be
S_{\text{c.t.}}=\int d^5x \left( \sqrt{\det G_{MN}}~ g^2 - 
\sqrt{\det G_{\mu\nu}}~[T_0\,\delta(y)+T_\pi\,\delta(y-\pi L)] \right).
\label{counterterms}
\ee
The four-dimensional effective Lagrangian including the radion
one-loop effective potential is then
\be
\mathcal L=-V+\pi L g^2 \phi^{-\frac{1}{3}}+
\frac{1}{2}(T_0+T_\pi)\phi^{-\frac{2}{3}}, 
\ee
where $V$ is the Casimir energy.
By considering $N_V$ vector multiplets and $N_h$ hypermultiplets
propagating in the bulk, the Casimir energy is~\cite{geromarianotoni}
\be
V\propto(2+N_V-N_h)\frac{1}{L^4 \phi^2}.
\label{Veffnomass}
\ee
For $2+N_V-N_h>0$ this gives a repulsive force.  Since $V'<0$ and
$V''>0$ one can generate a minimum at any desired value $\phi_0^{1/3}$
by choosing counterterms $g^2<0$ and $T_0+T_\pi<0$. This corresponds
to 5D de Sitter spacetime which is not consistent with supersymmetry.
For $2+N_V-N_h<0$ Eq.~(\ref{Veffnomass}) gives an attractive force and
no stable minimum can be created by adding counterterms.  The way out
is introducing an explicit mass scale in the
problem~\cite{Ponton:2001hq}. We will do this by introducing a
supersymmetric (odd) mass for some hypermultiplets that produces an
exponential localization of their lightest eigenstate on orbifold
fixed points.

\begin{figure}[hbt]
\centering
\psfrag{x}{$x$}
\psfrag{w}{$\omega$}
\epsfig{file=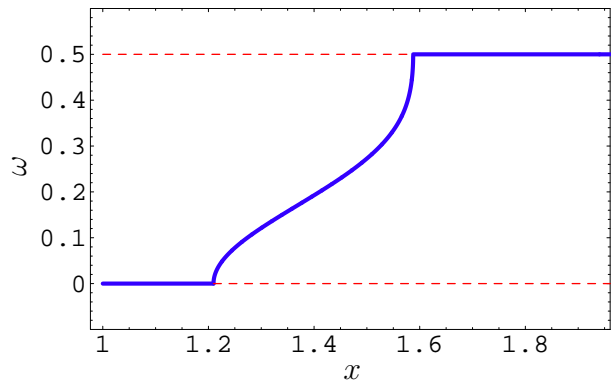}
\caption{ Solutions to the equation of motion for $\omega$. The blue
(solid) line corresponds to a minimum while the red (dashed) one is a
maximum.}
\label{fig1}
\end{figure}

If we consider $N_H$ hypermultiplets ``quasi-localized'' on one
brane by a common odd mass $M$, the effective potential can be cast
as~\cite{vonGersdorff:2003qf}
\be
V_{\text{eff}}=M^6L^2\frac{2+N_V-N_h}{64}\left(\frac{1}{x^6}f(\omega,x) 
+\frac{a}{x} + \frac{b}{x^2}\right),
\label{Veff}
\ee
where $x=ML\pi\phi^{1/3}$ and $V_{\rm eff}$ depends on $V_5$ only
through the flux $\omega$ as defined in Eq.~(\ref{wilsonflux}).  The
dimensionless counterterms $a$ and $b$ are related to the dimensionful
ones of Eq.~(\ref{counterterms}) as
\be
a=-\frac{64\pi^2 g^2}{(2+N_V-N_h)M^{5} },\qquad
b=\frac{32(T_0+T_\pi)\pi^2}{(2+N_V-N_h)M^{4}}.
\ee
The function
$f(\omega,x)$ has been computed in Ref.~\cite{vonGersdorff:2003qf} to be
\begin{eqnarray}
f(\omega,x)&=&-12\sum_{k>0}\frac{\sin^2(\pi \omega k)}{k^5}+
8\delta\!\intop_0^\infty dz\,z^3 
\log\left[1+\frac{(z^2+x^2)\sin^2(\pi\omega)}
{z^2\sinh^2(\sqrt{z^2+x^2})}\right]\nonumber \\
&\approx&-12\sum_{k>0}\frac{\sin^2(\pi \omega k)}{k^5}+
4\delta \sin^2(\omega\pi)F(x),
\label{approx}
\end{eqnarray}
where we have also defined $\delta=N_H/(2+N_V-N_h)$ and
\be
F(x)=e^{-2x}[3+6x+6x^2+4x^3].
\ee
Eq.~(\ref{approx}) is a good approximation for $x\apprge 1$.  We are
now looking for solutions $\omega=\omega(x)$ of the equations of
motion determined by (\ref{Veff}).  The solution(s) are plotted in
Fig.~\ref{fig1} for $\delta=1.5$ which clearly corresponds to
$2+N_V-N_h>0$ and $2+N_V-N_h-N_H<0$. We thus expect the following
behavior as a function of $M$: for $M=0$ ($x=0$) there is a minimum at
$\omega=0$ and a maximum at $\omega=\frac{1}{2}$ while for $M\to
\infty$ ($x\to \infty$) there is a minimum at $\omega=\frac{1}{2}$ and
a maximum at $\omega=0$ (simply because the contributions from the
massive hypermultiplets decouple as they become more and more
localized towards the branes). For intermediate values of $M$,
e.g.~$x_0\leq x\leq x_1$, there is a region where $\omega(x)\neq
0,\frac{1}{2}$ which corresponds to a minimum.
\begin{figure}[hbt]
\centering
\psfrag{x}{$x$}
\psfrag{V}{$V_{\rm eff}$}
\epsfig{file=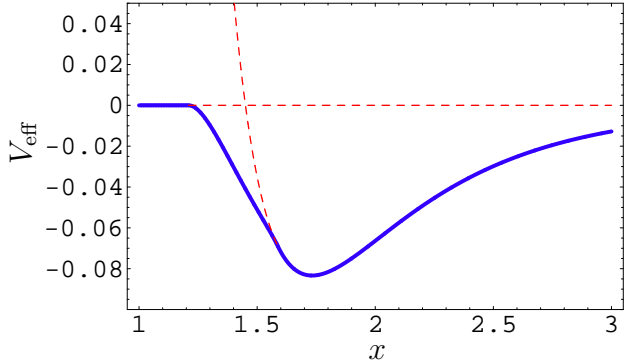}
\caption{ 
The Casimir energy for the different solutions of Fig.~\ref{fig1}.}
\label{fig2}
\end{figure}
The precise values of $x_0$ and $x_1$ can be determined in the
approximation (\ref{approx}) from
\be
\delta F(x_0) =3 \zeta(3),
\ee
\be
\delta F(x_1)=\frac{9}{4}\zeta(3).
\ee
Furthermore we can approximate the sum in Eq.~(\ref{approx}) by
e.g.~its first three terms which allows an analytic determination of
$\omega(x)$ in the region $x_0\leq x\leq x_1$
\be
\omega(x)=\frac{1}{2\pi}\arccos\left\{-\frac{81}{64}+\left[\left(\frac{81}{64}\right)^2-20+\frac{27}{4}\delta F(x)\right]^{\myfrac{1}{2}}\right\}.
\ee
The resulting Casimir energy $V(\omega(x),x)$ is plotted in Fig.~\ref{fig2}.
Notice that the minimum corresponds to $x>x_1$, i.e.~to
$\omega=\frac{1}{2}$. Also note that for $\delta\to 1$ the minimum
disappears: this is easily seen from the exact behaviour of $f(1/2,x)$ for
$x\to 0$, $x\to\infty$ which is given by
\be\label{cota}
\lim_{x\to 0} f(1/2,x)=-12 (1-\delta)\zeta(5),\qquad \lim_{x\to \infty} f(1/2,x)=-12\zeta(5).
\ee
This indicates that the function $x^{-6}f(1/2,x)$ only has a minimum for
$\delta>1$.

\begin{figure}[hbt]
\centering
\psfrag{x}{$x$}
\psfrag{V}{$V_{\rm eff}$}
\epsfig{file=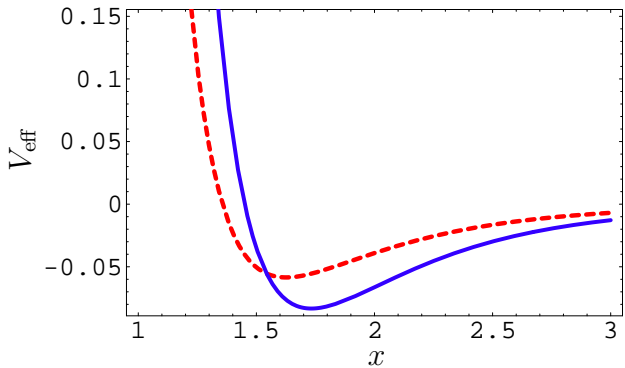}
\caption{ The Casimir energy for fixed $\omega=1/4$ (red/dashed line)
and $\omega=1/2$ (blue/full line).}  
\label{fig3}
\end{figure}

In the presence of supersymmetry breaking brane effects the value of
$\omega$ is fixed at tree level and the potential Eq.~(\ref{Veff})
(for $a=0$, $b=0$) has a minimum for any value of $\omega$.  In
Fig.~\ref{fig3} the Casimir energy is plotted for $\delta=1.5$ and
several values of $\omega$. Similar arguments to those following
Eq.~(\ref{cota}) also lead, for arbitrary $\omega$, to radion
stabilization only for $\delta>1$. Notice that for $x\geq x_1$,
i.e.~in the region of the minimum, the potential of Fig.~\ref{fig2}
coincides with that of Fig.~\ref{fig3} for $\omega=\frac{1}{2}$. This
means that for all practical purposes we can consider the potential
Eq.~(\ref{Veff}) with $\omega$ constant (i.e.~not depending on $x$) and
the case of no supersymmetry breaking brane effects just corresponds
to $\omega=\frac{1}{2}$.

\begin{figure}[hbt]
\centering
\psfrag{aaa}{$x$}
\psfrag{bbb}{$V_{\rm eff}$}
\epsfig{file=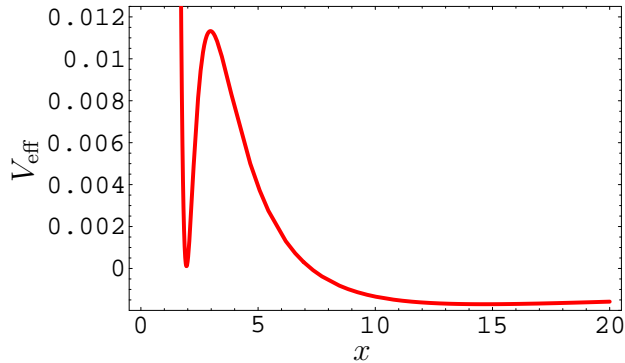}
\caption{The effective potential corresponding to the $\omega=1/2$ case with
fine-tuned counterterms.}  
\label{fig4}
\end{figure}

The potential corresponding to the Casimir energy should then be completed
with the counterterm Lagrangian with $g^2>0$ and $T_0,\,T_\pi>0$ to
fine-tune to zero the 4D cosmological constant. In particular, the
condition for supersymmetric AdS$_5$ space is~\cite{Bagger:2002rw}
\be
T_{0,\pi}\leq \sqrt 6 g M_5^{3/2},
\ee
which translates in our case to
\be
\frac{1}{4\pi}\frac{M}{M_5}
\leq \left[\frac{6 |a|}{(2+N_V-N_h)\pi b^2}\right]^{1/3}.
\ee
When we include the counterterms $a<0$ and $b>0$ (fine-tuned to have a
zero cosmological constant) the effective potential develops an extra
AdS$_4$ minimum and goes to zero from below for $x\to\infty$. An
example is shown in Fig.~\ref{fig4} where the effective potential
corresponding to $\omega=1/2$ and $a=-0.05$ is presented.
In any case we should be concerned for the tunneling from the
Minkowski to the AdS vacuum. In the rest of this section we will
estimate the tunneling rate and find it to be exceedingly small so that
our vacuum with zero cosmological constant is stable. 

A priori, the quantum mechanical stability of the false Minkowski
vacuum state is not guaranteed since the decay of the false vacuum
into the AdS vacuum may proceed by quantum tunneling in a finite sized
bubble.  In order to estimate the tunneling probability, we use a
dimensional argument~\footnote{We also have performed a detailed
computation of the tunneling rate following
Ref.~\cite{Coleman:1980aw}. It reproduces the result obtained from the
dimensional argument.}.  The kinetic term for the field $\phi$ assumes
the form

\begin{equation}
\label{kin}
{\cal L}_{\rm kin}=\frac{1}{6}M_4^2\frac{\partial_\mu\phi\partial^\mu
\phi}{\phi^2},
\end{equation}
where $M_4$ is the four-dimensional Planck scale. The potential for
the field $\phi$ can be generically written in the form $M^4{\cal
V}(\phi)$, where ${\cal V}(\phi)$ is a dimensionless quantity.
The equation for the flat space instanton which controls
the tunneling rate from the false to the true vacuum is given by

\begin{equation}
\label{mm}
\partial_\rho^2 \phi_{\rm inst}+\frac{3}{\rho}\partial_\rho \phi_{\rm inst}-2
\frac{\left(\partial_\rho \phi_{\rm inst}\right)^2}{\phi_{\rm inst}}
=6\phi_{\rm inst}^2\frac{M^4}{M_4^2}
\left.\frac{d{\cal V}}{d\phi}\right|_{\phi=\phi_{\rm inst}},
\end{equation}
where $\phi_{\rm inst}=\phi_{\rm inst}(\rho)$ is the instanton
solution. It is a function of the variable $\rho=\sqrt{\vec{x}^2
+t^2}$ which makes the $SO(4)$ symmetry of instanton solutions
at zero temperature manifest.  The solution to Eq.~(\ref{mm}) will have the
form

\begin{equation}
\label{o}
\phi_{\rm inst}=f\left(z\frac{M_4}{M^2}\right),
\end{equation}
where $z$ is a dimensionless variable. Eq.~(\ref{o}) tells us that the
critical size $\rho_c$ of the bubble by which tunneling may proceed is
of the order of

\begin{equation}
\rho_c\sim \frac{M_4}{M^2}.
\end{equation}
As a consequence it is easy to see that the instanton action is of the
order of
\begin{equation}
S_{\rm inst}\sim \rho_c^4\, M^4 {\cal V}\sim \left(\frac{M_4}{M}\right)^4.
\label{instantonaction}
\end{equation}
The corresponding vacuum tunneling probability per unit space-time
volume is of the order of
\begin{equation}
{\cal P}\sim e^{-S_{\rm inst}}\sim e^{-10^{60}},
\end{equation}
where in the last passage we have chosen $M\sim$ 1 TeV. This is so small
that the false vacuum is essentially stable on cosmological times. 

At this stage the reader might wonder why the tunneling probability is
so small and be puzzled by the fact that decreasing the size of the
barrier separating the false from the true vacuum by lowering $M$, the
tunneling probability drops exponentially.  The reason is the
following: in terms of the field $\sigma= \frac{M_4}{\sqrt{3}}\,{\rm
ln}\,\phi$ which canonically normalizes the kinetic term (\ref{kin}),
the height of the barrier separating the two vacua is of the order of
$M^4$, while the two vacua are far from each other at a distance $\sim
M_4\gg M$ in field space. Despite the small barrier between the two
vacua, in order to tunnel a macroscopic bubble has to be
nucleated. Indeed, the smaller the barrier the bigger the size of the
bubble, $\rho_c\sim \frac{M_4}{M^2}$. This amounts to saying that,
despite the small energy-density $\sim M^4$ inside the bubble, its
radius is so large that the total energy cost is measured by $M^4
\rho_c^4$ which is precisely of the order of $S_{\rm inst}$ in
Eq.~(\ref{instantonaction}).

\section{\sc Conclusion and Outlook}

In this paper we have addressed the issue of radion stabilization in
supersymmetric five-dimensional theories compactified on the orbifold
$S^1/\mathbb{Z}_2$ where supersymmetry is broken by the Scherk-Schwarz
mechanism.  SS supersymmetry breaking can be straightforwardly
implemented in the off-shell version of 5D supergravity using tensor-
or linear multiplets as compensators. We have shown in detail that
both formulations are equivalent both locally and globally and that a
one loop analysis is needed to fix the supersymmetry breaking flux for
the auxiliary gauge field of the $U(1)_R$ automorphism symmetry.  In a
globally inequivalent version, the linear multiplet formulation allows
for a tree level breaking of supersymmetry~\cite{Rattazzi:2003rj}. We
have shown that the bulk radius may be stabilized in the presence of a
number $N_H$ of quasi-localized bulk fields whose contribution to the
one-loop Casimir energy is such that, once introduced a bulk
cosmological constant and brane tensions to achieve zero
four-dimensional cosmological constant, the radion field is stabilized
in a metastable Minkowski vacuum.  For the mechanism to work we have
found a lower bound on $N_H$ given by $2+N_V+N_h$ where $N_V$ $(N_h)$
is the number of massless gauge multiplets (hypermultiplets)
propagating in the bulk. We have shown that the probability of
decaying from such a vacuum to the true one with negative cosmological
constant is completely negligible.

In the metastable vacuum the squared mass of the canonically
normalized radion field is given by $\sim$ (one-loop factor)$\times
\frac{M^4}{M_4^2}$. Since the size of the odd-mass term $M$ may be
taken to be of the order of 10 TeV, we conclude that the radion field
acquires in the metastable vacuum a mass around $(10^{-3}-10^{-2})$
eV. This range of masses is interesting for present and future
measureaments of deviations from the gravitational inverse-square law
in the millimeter range~\cite{nelson}.

We should also comment at this point about the relationship of radion
stabilization with the hierarchy problem. Unlike in those approaches
where a warped geometry solves the hierarchy problem, in flat space we
must invoke supersymmetry for solving it. Therefore even if solving
the hierarchy problem by the radion stabilization in warped geometries
was a real issue, here it is not such. Our only concern was to obtain
a physical radius $\simlt 1/$TeV. However this range is technically
natural since we are introducing bulk masses in the TeV range. A
different (not unrelated) issue is the origin of the weakness of
gravitational interactions in the 4D theory and its relation with
radion fixing. Here we have been working in a 5D gravity theory, with
a 1/TeV length radius, and therefore the presence of submillimeter
dimensions is not consistent with our mechanism for radion
stabilization. On the other hand the relation between the Planck
scales in the 4D and 5D theories, $M_4^2=M_5^3R$, with $R\sim 1/$TeV
implies that the scale where gravity becomes strong in the 5D theory
is much higher than $1/R$. This means that gauge interactions of the 5D
theory become non-perturbative at a scale $M_s\ll M_5$ in the
multi-TeV range. The theory should then have a cutoff at the scale
$M_s$ where a more fundamental theory should be valid. An example of
such behaviour is provided by Little String Theories (LST) at the
TeV~\cite{LST,Ant} where the string coupling $g_s\ll 1$ and $M_s$ and
$M_5$ are related by $M_5^3=M_s^3/g_s^2$. In other words $M_5$ does no
longer play the role of a fundamental field theoretical cutoff
scale. In these theories the weakness of the gravitational
interactions is provided by the smallness of the string
coupling. Moreover a class of LST has been found~\cite{Ant} where the
Yang-Mills coupling is not provided by the string coupling but by the
geometry of the compactified space where gauge interactions are
localized, e.g.~$g_{YM}\sim \ell_s/R$. Since the field theory has a
cutoff at $M_s$ the consistency of the whole picture relies on the
assumption that there is a wide enough range where the 5D field theory
description is valid.

Finally, we should also be concerned about the backreaction of the
Casimir energy and the counterterms on the originally flat 5D
gravitational background.  A dimensional analysis shows that the
effect of the counterterms by themselves would result in a warp factor
with a functional dependence on the extra coordinate as $a(\epsilon
My)$, where $\epsilon=\mathcal O(M/M_5)^{3/2}\equiv \mathcal
O(M/M_4)\sim 10^{-15}$ for $M\sim$ TeV.  Such a warping is competely
negligible. One can also show that the size of the gravitino bulk and
brane masses generated by the counterterms are of the order of the
radion mass and thus negligible as compared to the size of
supersymmetry breaking contributions.

\vspace*{7mm}
\subsection*{\sc Acknowledgments}

\noindent This work was supported in part by the RTN European Programs
HPRN-CT-2000-00148 and HPRN-CT-2000-00152, and by CICYT, Spain, under
contracts FPA 2001-1806 and FPA 2002-00748. The work of one of us
(G.v.G.) is supported by the DAAD. Another of us (A.R.) would like to
thank the Theory Department of IFAE, where part of this work has been
done, for hospitality. We would like to thank J.~Garriga and
O.~Pujolas for discussions.


\end{document}